\documentclass[aps,pre,twocolumn,superscriptaddress]{revtex4-2}
\usepackage{graphicx}
\usepackage{hyperref}
\usepackage{color}
\usepackage{siunitx}
\DeclareSIUnit\a{\textit{a}} % dimensionless length unit
\DeclareSIUnit\cpa{\textit{c}/\textit{a}} % dimensionless frequency

\begin{document} 

%\preprint{}


\title{Stealthy and hyperuniform isotropic photonic bandgap structure in 3D}% Force line breaks with \\


%\author{}
%\thanks{}
%\affiliation{}


\author{Lukas Siedentop}
\affiliation{Department of Physics, University of Konstanz, 78457 Konstanz, Germany}
% \altaffiliation[Also at ]{Physics Department, XYZ University.}%

\author{Gianluc Lui}
\affiliation{Advanced Technology Institute and Department of Physics,
University of Surrey, Guildford, Surrey GU2 7XH, United Kingdom}

\author{Georg Maret}
\affiliation{Department of Physics, University of Konstanz, 78457 Konstanz, Germany}

%\author{Antonio M. Puertas}
%\affiliation{Department of Applied Physics, University of Almeria, 04120 Almeria, Spain}

\author{Paul M. Chaikin}
\affiliation{Department of Physics, New York University, New York, NY 20012 USA}

\author{Paul J. Steinhardt}
\affiliation{Department of Physics, Princeton University, Princeton, NJ 08544 USA}

\author{Salvatore Torquato}
\affiliation{Department of Physics, Princeton University, Princeton, NJ 08544 USA}
\affiliation{Department of Chemistry, Princeton University, Princeton, NJ 08544, USA}
\affiliation{Princeton Materials Institute, Princeton University, Princeton, NJ 08544, USA}
\affiliation{Program in Applied and Computational Mathematics, Princeton University, Princeton, NJ 08544,
USA}

\author{Peter Keim}%
\thanks{corresponging author: peter.keim@ds.mpg.de}%
\affiliation{Max-Planck-Institute for Dynamics and Self-Organization, 37077 G\"ottingen, Germany}
\affiliation{Institute for the Dynamics of Complex Systems, University of G\"ottingen, 37077 G\"ottingen, Germany}

\author{Marian Florescu}
\affiliation{Advanced Technology Institute and Department of Physics,
University of Surrey, Guildford, Surrey GU2 7XH, United Kingdom}

\date{\today}% It is always \today, today,

\begin{abstract}

In photonic crystals the propagation of light is governed by their photonic band structure, an ensemble of propagating states grouped into bands, separated by photonic band gaps. Due to discrete symmetries in spatially strictly periodic dielectric structures their photonic band structure is intrinsically anisotropic. However, for many applications, such as manufacturing  artificial structural color materials or developing photonic computing devices, but also for the fundamental understanding of light-matter interactions, it is of major interest to seek materials with long range non-periodic dielectric structures which allow the formation of {\it isotropic} photonic band gaps. Here, we report the first ever 3D isotropic photonic band gap for an optimized disordered stealthy hyperuniform structure for microwaves. The transmission spectra are directly compared to a diamond pattern and an amorphous structure with similar node density. The band structure is measured experimentally for all three microwave structures, manufactured by 3D-Laser-printing for meta-materials with refractive index up to $n=2.1$. Results agree well with finite-difference-time-domain numerical investigations and a priori calculations of the band-gap for the hyperuniform structure: the diamond structure shows gaps but being anisotropic as expected, the stealthy hyperuniform pattern shows an isotropic gap of very similar magnitude, while the amorphous structure does not show a gap at all. The centimeter scaled microwave structures may serve as prototypes for micrometer scaled structures with bandgaps in the technologically very interesting region of infrared (IR).
\end{abstract}

% insert suggested keywords - APS  authors don't need to do this
%\keywords{}

\maketitle

\section{Introduction}

The manipulation of light propagation by employing periodic dielectric structures is widely used in technology, e.g in dielectric mirrors and anti-reflection coatings. For this purpose, 1D periodic structures reflect or transmit only a narrow part of the electromagnetic spectrum due to Bragg-scattering. Bragg-scattering strongly depends on the orientation of the structure with respect to the incident wave and is thus intrinsically anisotropic. The generalization to 3D structures leads to so called photonic crystals with stop bands for the propagation of light in various directions of the given Brillouin zone in close analogy to electronic band gap formation~\cite{John1987, Yablonovitch1987, Joannopoulos1997}. Here, the typical length scale is given by the Bragg condition, thus the dielectric meta-material is structured on a scale comparable to the electromagnetic wavelength: centimeter range for microwaves and sub micron range for visible light.\\

\begin{figure*}
    \centering
    \includegraphics[width=\textwidth]{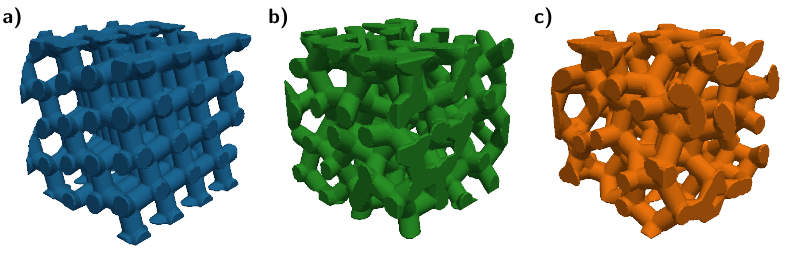}
    \caption{Excerpts of the three cylinder model structures under investigation. a): a diamond pattern, possessing a well known but anisotropic band structure b): the optimized stealthy hyperuniform pattern with isotropic band structure, c): an isotropic pattern generated from an amorphous glassy seed pattern from Monte-Carlo simulations for comparison.
%with the protocol explained in \cite{Florescu2009} adapted to 3D.
    }\label{fig:ch4_networkcomp_dia_hpu_gla}
\end{figure*}

Two-dimensional photonic structures exhibiting a complete photonic band gap, for both TE and TM polarizations, are rather challenging to realize. This is because, even for periodic structures, the architectures needed are rather different, TM-polarization photonic band gap opening very easily in isolated scatterer architectures, whereas the optimal favoured  architecture for opening of TE-polarization band gaps consists of connected dielectric network structures. Complete bandgaps in 2D can be opened in structures which reach a  compromise between the two architectures and consist of dielectric scatterers connected by narrow dielectric veins \cite{Joannopoulos2008, Fu2005, Florescu2009} but the largest complete gaps reach a rather modest size of just 15\% of the midgap frequency for silicon-air index of refraction contrast \cite{Fu2005}. In contrast, three dimensional photonic structures based on a diamond-network architecture are naturally adapted for the opening of complete photonic band gaps and can reach band gaps of about 30\% of the midgap frequency for the same  index of refraction contrast \cite{Yablonovitch1989}.  However, as is the case with all periodic structures, the high-symmetry directions in the underlying FCC lattice, induces strongly anisotropic photonic band gaps.\\

For technological applications but also from a fundamental point of view it is of enormous interest to find materials with isotropic photonic band gaps where the photonic density of states disappears in all directions. It has been long argued that isotropic band gaps will form in dielectric meta-materials whose structure is itself isotropic~\cite{Man2005}. So called hyperuniform structures~\cite{Florescu2009}, where the structure factor vanishes in the long-wavelength limit were tested to have a band gap in 2D for microwaves ~\cite{Man2013, Man2013a, Aubry2020}. For a $D$-dimensional point pattern in a $D$-dimensional spherical sampling window with radius $R$, hyperuniformity is defined as the number variance of points contained within the spherical window, $\sigma (R)$ when averaged over all possible positions of the window within the domain being considered. The point pattern is hyperuniform if $\sigma(R)$ grows as $R^{D-1}$; that is the number variance is proportional to the surface area of the sampling window rather than its volume as is the case of e.g. Poisson point patterns \cite{Torquato2003, Batten2008}. Hyperuniform point patterns include all photonic crystals, quasicrystals and a subset of disordered structures. They possess zero density fluctuations on infinite length scales within the structure so their structure factor $S(k)$ vanishes for $k\to 0$.\\

\begin{figure*}
    \centering
    \includegraphics[width=.9\textwidth]{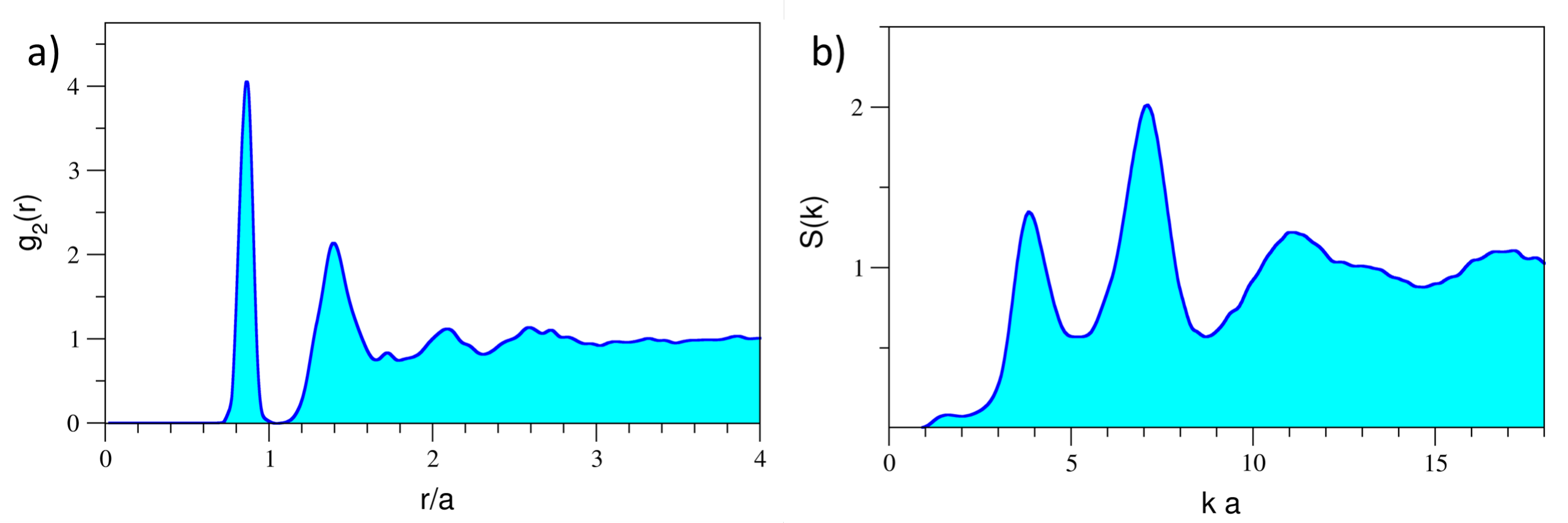}
    \caption{Two-point correlation function (a) and structure factor (b) for the optimized stealthy hyperuniform point pattern constructed from an $N=1000$ CRN model, with $\chi=0.03$. For stealthyness, the structure factor is forced to zero for $ka<1$.
    }\label{fig_g_S}
\end{figure*}
\begin{figure*}
    \centering
    \includegraphics[width=.9\textwidth]{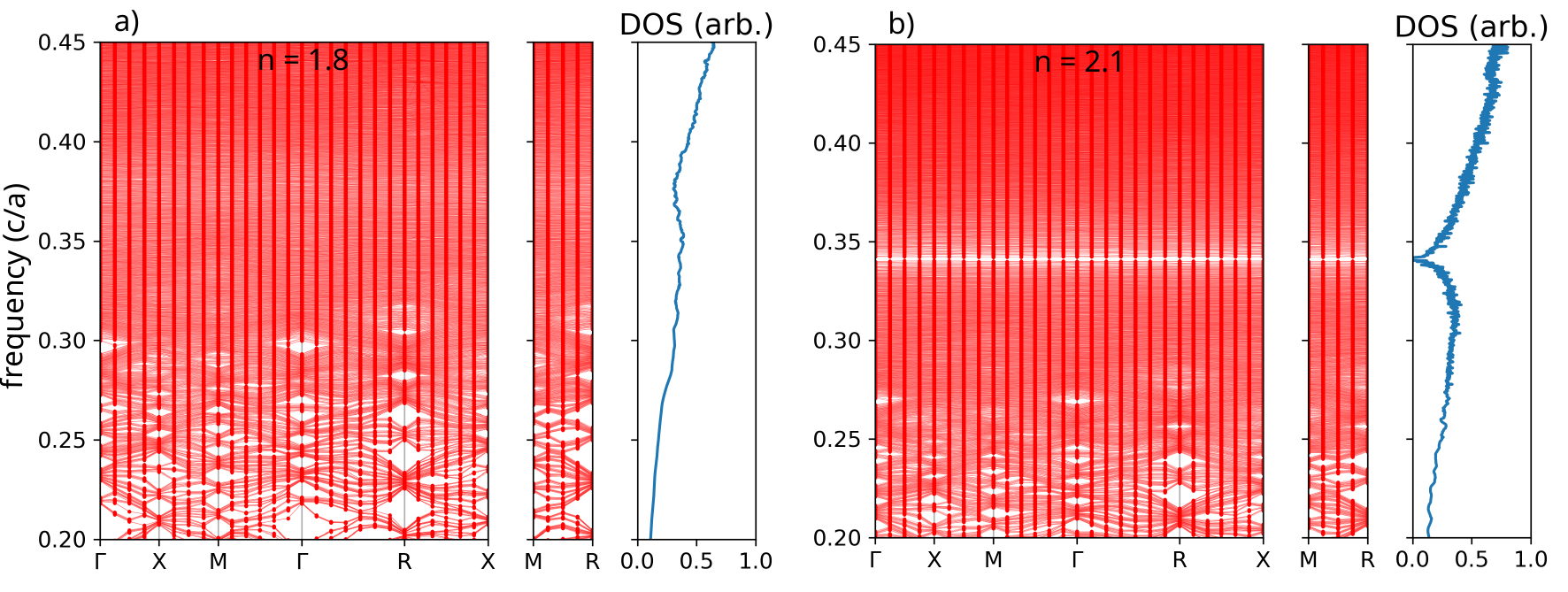}
    \caption{Band structure for dielectric-rod decorated optimised hyperuniform structures for contrast of the indices of refraction corresponding to (a) composite of alumina $n=1.8$, and (b) composite of titania $n=2.1$. While for $n=1.8$ only a dip in the DOS is visible, a small but complete and isotropic band gap opens up for $n=2.1$. The band structure and density of states are calculated for a sample hyperuniform network of 1000 vertices decorated with dielectric rods of radius $r/a=0.3$ contained in a $[10\times a]^3$-supercell.
    }\label{fig_bs_dos}
\end{figure*}

For relatively large refractive indices, Muller~et.~al. and Aeby~et~al. have demonstrated 3D emergent isotropic band gaps in the near infrared~\cite{Muller2014, Muller2017, Aeby2022}. Their structures are based on disordered jammed packings that are hyperuniform and nearly stealthy. Beside hyperuniformity, the optimization of short range order to tailor Bragg scattering at the Brillouin zone is of key importance, too~\cite{Liew2011, FroufePerez2016}. Furthermore, internal Mie-resonances within the high-index material (spheres or cylinders) affect the photonic density of states~\cite{Rockstuhl2006, Rockstuhl2009, Schertel2019}. To characterize the short range order, Sellers~et~al. introduced the concept of local self-uniformity in cylinder based structures, where Mie-resonances within the cylinders have to interfere constructively with the structural arrangement of the cylinders~\cite{Sellers2017}. 3D Amorphous structures with diamond-like local tetrahedral order were investigated numerically \cite{Edagawa2008} and experimentally in the microwave regime \cite{Imagawa2010} showing photonic band gaps which were compared to diamond structures. It is important to stress the differences in the size of the PBG when comparing disordered non-stealthy hyperuniform structures to their stealthy hyperuniform counterparts as the system size increases. A comprehensive 2D study of disordered structures that range from nonhyperuniform to standard hyperuniform and stealthy hyperuniform ones revealed that the apparent PBGs rapidly close as the system size increases. This is for all disordered networks under consideration, except for the stealthy hyperuniform structures where the PBG persists \cite{Klatt2022}. For the same reasons, we expect that 3D stealthy hyperuniform dielectric networks have such PBG superiority.\\

In the present work, we investigate three photonic structures composed of an interconnected tetravalent network of cylinders a) of an anisotopic diamond lattice, b) an optimized isotropic and stealthy hyperuniform structure, and c) an isotropic network structure constructed from a glassy, random hard sphere packing seed pattern obtained from computer simulations. \autoref{fig:ch4_networkcomp_dia_hpu_gla} visualizes the three structures with the diamond-structure in blue, the stelathy hyperuniform pattern in green and the disordered pattern in red. For the diamond structure (blue) we expect a band structure but being strongly directional. The isotropic glass structure (red) is constructed for comparison and as reference for the optimized hyperuniform structure (green). In analogy with periodic structures, we define a length scale $a=L/\sqrt[3]{N}$ such that an N-point pattern in a cubic box of side length $L$  has a scatterer density of $1/a^3$. Samples are shown in a cube of $L = 4 \times a$ side length, while the investigated structures in the spectrometer have a side length of $L = 10 \times a$ .\\

\begin{figure*}
    \centering
    \includegraphics[width=0.65\textwidth]{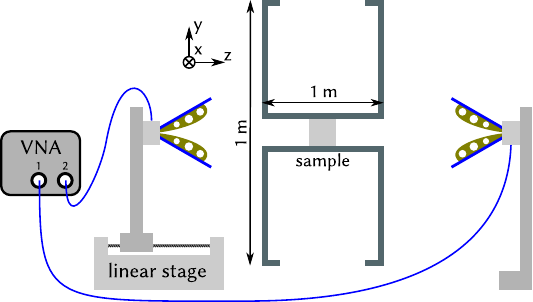}
    \caption{Sketch of the setup to measure the transmittance through macroscopic photonic structures. It consists of a vector network analyzer (VNA), two horn antennas, and a $1~\mathrm{m}$ wave-guide with a $10~\mathrm{cm}$ quadratic aperture in a $1~\mathrm{m}^2$ beam block. The samples are put in the middle of the waveguide. The position of the left antenna relative to the waveguide can be adjusted with a linear stage in the range of \SIrange{0.1}{0.38}{\m}, the distance of the right antenna to the waveguide is \SI{0.4}{\m}. \num{100} measurements are averaged at different positions of the left antenna.}\label{fig:experimental_setup}
\end{figure*}

The optimized hyperuniform structure was constructed by combining continuous random networks (CRN) inspired by models of amorphous silicon \cite{Barkema2000, Edagawa2008, Xie2013} and hyperuniformity concepts \cite{Torquato2003, Batten2008, Florescu2009, Sellers2017}. The CRN structures are generated by annealing a completely random four-fold coordinated network using the Wooten-Winer-Weaire (WWW) algorithm \cite{Weaire1971a, Weaire1971b}. The algorithm proceeds by introducing coordination-preserving Stone-Wales defects at random positions in the structure followed by subsequent relaxation of the structure with a Keating potential to minimize the spread in the distributions of the next-neighbour particle distance and the angles made by the tetrahedral bonds in the network (typical values for standard deviations of the distributions for optimised CRN are around $\sigma_{d}\approx 5\% $ and $\sigma_\theta\approx 9\%$) \cite{Hejna2013, Sellers2017}. While the continuous random network structures obtained through the Wooten-Winer-Weaire algorithm present well-defined short-range order, they are not yet hyperuniform. For photonic applications \cite{Florescu2009}, we are interested in an optimized subcategory of hyperuniform structures, namely “stealthy” hyperuniform structures \cite{Torquato2003, Batten2008}.\\

For stealthy hyperuniform point patterns, the structure factor $S(k)$ is statistically equal to zero for a finite range of wave numbers smaller than a certain critical wave vector $k_C$, i.e., $S(k < k_C ) = 0$. The stealthiness parameter $\chi= M/3N $ is defined as the ratio between the number of $k$ vectors for which the structure factor is constrained to vanish, $M$, and the total number of $k$ vectors associated with the pattern, $3N$ (with $N$ the number of points in the pattern). Here, we employ continuous random networks generated by using the WWW algorithm; the structures are subsequently made hyperuniform by forcing the structure factor at a fixed number of $k$ values to vanish. To maintain the well-defined short-range order associated with the CRN we employ only a small number of wavenumbers. In Fig. \ref{fig_g_S}, we present the two-point correlation function, $g_2(r)$ and the structure factor $S(k)$, for an $N=1000$ optimised stealthy hyperuniform pattern. Here, we enforce the structure factor to vanish for the smallest 100 $k$ values around the origin (i.e. $\chi = 100/3000\approx 0.03$). In Fig. \ref{fig_bs_dos}, we present the corresponding band structure and density of states calculations for a structure built by decorating the hyperuniform point pattern with dielectric rods of various indices of refraction. The original stealthy hyperuniform point pattern is generated under periodic boundary conditions, and the band structure was calculated using a supercell approximation. The refractive indices were chosen to map the experimental accessible ones. While for $n=1.8$, only a dip in the density of states around $f=0.75$ (in units of $c/a$) is visible, for $n=2.1$ a complete and isotropic gap opens up around $f=0.34$. The width of the gap is $0.6\%$. We note that for higher refractive indices, e.g. $n=3.4$ the gap width increases up to $14\%$ and is centered at $f=0.24 c/a$ (not shown here). As shown in Fig. \ref{fig_bs_dos}b), the band-gap is independent of the orientation of the wave-vectors and it is isotropic as intended. 

\subsection{Experimental and Numeric Results}

Macroscopic structures given by cubes with \SI{100}{\mm} side-length and operational wavelengths of about \SI{30}{\mm} where manufactured, corresponding to frequencies in the \SI{10}{\GHz} range. This is inspired by the seminal work of Yablonovitch~\cite{Yablonovitch1989} and previous applications to photonic amorphous diamonds~\cite{Edagawa2008,Imagawa2010}. All three structures where realized with a 3D printer using selective laser sintering (SLS) of a compound material. This compound material consists of a polymer (Nylon) with additives of high refractive index materials as Al$_2$O$_3$ and TiO$_2$. The mass ratio of the oxides in Nylon were increased up to a value until SLS printing failed to produce a mechanical stable object. For Al$_2$O$_3$, a refractive index of about $n=1.8$ was accessible, while for TiO$_2$ we manufactured samples with $n=2.1$ in the given frequency range of the micro-waves. Details on the printing procedure are given in \autoref{sec:ch3_sintratec}.\\

\begin{figure*}
    % Figure Disukussion:
    \centering
    \includegraphics[width=.91\textwidth]{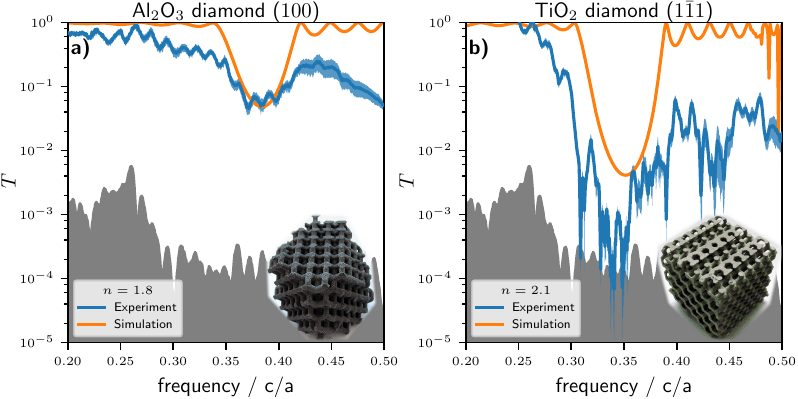}
    \caption{Results of transmission measurements for two macroscopic realizations of the meta-material of diamond. Spectra are taken in two different directions of the lattice and for two different refractive indices $n=1.8$ and $n=2.1$. The cubic samples have a side-length of \SI{10}{\cm}. The Al$_2$O$_3$ samples suffer from mechanical damage due to their brittleness, as exemplary shown in the inset image of the diamond in a). The peak position as well as the depth of the peak coincide well for both refractive indices and orientations of the diamond lattice.}\label{fig:experimental_diamond}
\end{figure*}

The transmittance of the samples is measured by placing the sample in a wave guide that just fits the sample, as sketched in \autoref{fig:experimental_setup}. Two antennas emit respectively receive linear polarized electromagnetic radiation from \SIrange{0.5}{13.5}{\GHz}, as controlled by a vector network analyzer. More details about the setup are given in \autoref{sec:method_macrospectroscopy}. Images of the laser-sintered diamond structures are shown in the insets of \autoref{fig:experimental_diamond} along with the measured transmittance curves (blue) and the numerically calculated ones (orange) using the software package \texttt{meep}, as described in \autoref{sec:ch3_FDTD}. The diamond structure made of Al$_2$O$_3$ is investigated in (100) direction while the TiO$_2$ is analysed in (111) direction. Both structures acts as proof of principle for the experimental setup. The positions and depth of the gaps coincide very well in both directions for the given refractive indices. Note, that for the (111) direction with $n=2.1$, the transmission is reduced by almost three orders of magnitude at a frequency of $f=0.34 c/a$. The experimental transmission (blue) is smaller compared to numerical data (orange), since a) it contains some absorption which is absent in the simulation and b) the antennas naturally only transmit (or receive) within a finite angle. Thus losses are expected via radiation beside the "optical" axis. This is different in the simulation, where the flux planes detects radiation in and from all directions and metallic boundary conditions are applied perpendicular to the optical axis.\\ 

The dark spectrum of the setup is shown as the grey area at the bottom of each spectrum. It is the experimentally minimal detectable signal of the given device. It reflects the dynamic range of the vector network analyzer, the efficiency of the antenna and the mode-structure of the wave-guide. As function of frequency, the dynamic range of the setup is determined to be $30 - 40 \mathrm{dB}$. To gain insight about the sensibility of the device with respect to the local geometry, one antenna was mounted on a linear stage (along the optical axes) and data were taken at about 100 different positions of the antenna. The shaded blue region of the experimental data denotes the standard deviation from averaging about this 100 positions. As can be seen in \autoref{fig:experimental_diamond}, the measurement is rather stable with regards to linear translation of the antenna. However, beside the gap in transmission, the experimental data hardly recover the maximal transmission of $T=1$, especially on the high-frequency side. We attribute this to scattering beside the optical axis (as in the gap) which is stronger for high frequencies. While FDTD-simulations measures the hemispherical reflection, the experiment is sensitive to directional reflection, see (\autoref{sec:ch3_FDTD}). This feature, together with depolarisation (not captured by the antenna) explains the reduced transmission of experimental data versus FDTD-simulations.\\

\begin{figure*}
    % Figure Disukussion:
    \centering
    \includegraphics[width=.91\textwidth]{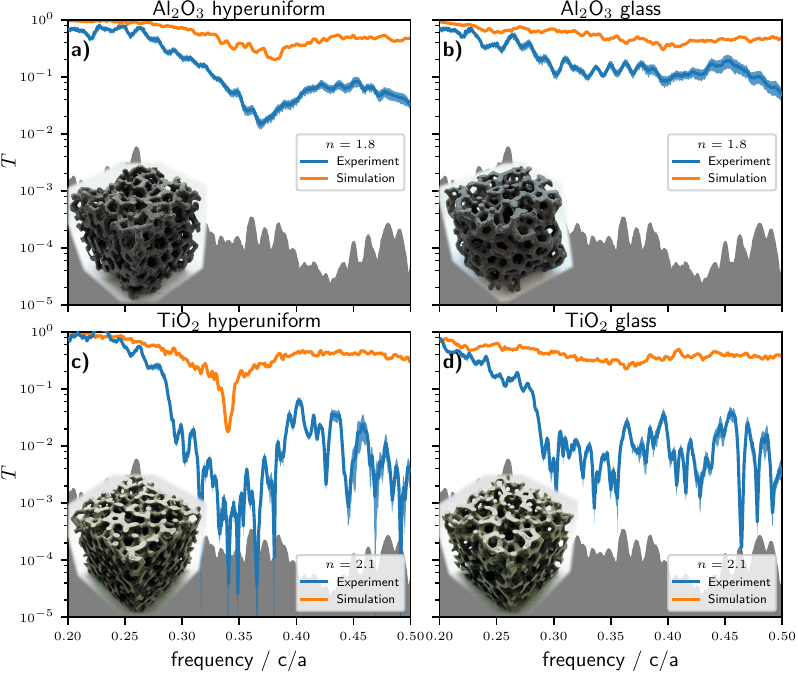}\\
    \caption{Results of the macroscopic realizations for two different refractive indices $n=1.8$ (upper row) and $n=2.1$ (lower row) for the stealthy hyperuniform structure (left column) and the amorphous sample constructed from structural glass data (right column). The cubic samples in experiment have a side-length of $10~\mathrm{cm} = 10a.$ }\label{fig:experimental_hpu_glass}
\end{figure*}

The structure of interest is the stealthy hyperuniform one, shown in the left column of \autoref{fig:experimental_hpu_glass}. For $n=1.8$ (upper row) a dip at $f= 0.37$ shows up while at $n=2.1$ (lower row) the gap is positioned at $f=0.34$ and is $20 \mathrm{dB}$ for the FDTD simulations and even $30 \mathrm{dB}$ for the experimental data. Here, the shaded blue region not only averages about $100$ different positions of the linear stage along the optical axis but the tree different orientations of the cubes. Within the error bars we did neither observe differences due to the orientations with respect to the optical axis - as intended by isotropy and numerically proven in Fig. \ref{fig_bs_dos}b) - nor due to differences of the orientation with respect to the polarization of the antenna. Since the reduction in transmission of the experimental data are significantly stronger in the disordered isotropic hyperuniform structure compared to the diamond lattice, it is less due to absorption as due to scattering. This is also seen in the transmission-data from the reference structure constructed from a 3D glass (right column), which completely lacks a photonic band gap, but a significant reduction of the experimental spectrum compared to FDTD simulation is visible (with stronger enhancement for the larger refractive index). While both structures are tetra-valent and look similar by eye, the standard deviation of the node distance varies significantly: it is \SI{0.035}{\a} for the hyperuniform pattern and \SI{0.09}{\a} for the glass pattern. The same holds for the standard deviation of angle distribution of the network which is \SI{9}{\degree} of the stealthy-hyperuniform structure and \SI{14}{\degree} for the glass pattern. The optimised stealthy and hyperuniform pattern allows for destructive interference but---unlike in periodic structures---in all spatial directions.

\section{Methods}
\label{methods}

\subsection{Sample production with 3D printing}
\label{sec:ch3_sintratec}

Selective laser sinthering (SLS) offers the possibility to construct samples from in principle any powder (partially) composed of meltable grains. The only restriction is the interconnected morphology of the structure, providing the chance to produce photonic structures of any kind. The refractive index of the structure is tunable by carefully choosing the powder constituents. The 3D printer used in this work is a Sintratec Kit~\cite{Knapp2016}. The meltable grains are made of Nylon (PA12). They are around \SI{50}{\um} in size and the melting point of Nylon is approximately \SI{185}{\celsius}~\cite{Knapp2016}. Those grains were mixed with grains of (Al$_2$O$_3$) or (TiO$_2$) to increase the refractive index of the compound material above that one of PA12. Samples with a maximum sidelength of \SI{100}{\mm} can be realized with this specific machine.\\

Printing protocols of powders with the two different additives aluminum oxide (Al$_2$O$_3$) and titanium dioxide (TiO$_2$) are developed similar to the procedure described in references~\cite{Imagawa2010, Edagawa2013}. The aim is to print structures of a wide range of refractive indices. The base material Nylon (PA12), (C$_{12}$H$_23$NO)$_n$) has a refractive index $n_\text{PA12} \approx \num{1.5}$ in the frequency regime of interest \SIrange{0.5}{18}{\GHz}. Undyed PA12 (AdSint PA12 L nat, Advanc3D Materials GmbH) with about \SI{40}{\um} grain size darkened with a small amount of carbon black to aid laser absorption is used for Al$_2$O$_3$ powder mixtures and grey material from Sintratec with about \SI{50}{\um} grain size is used for the TiO$_2$ mixtures. All materials have a small loss tangent~\cite{Imagawa2010} in the \si{\GHz}-regime thus absorption is neglected in the accompanying numeric investigations. Al$_2$O$_3$ has $n_\text{Al$_2$O$_3$} \approx 3$~\cite{Vila1998} and material with a grain size  of \SI{50}{\um} is used (product number 007-0160, Final Advanced Materials). TiO$_2$ has an refractive index of $n_\text{TiO$_2$} \approx 10$~\cite{Ahmad2012, Wypych2014} in the bulk. It is found that the commonly available nanosized TiO$_2$ of particle sizes around \SI{500}{\nm} tends to form agglomerates that are difficult to compound with the PA12. It is also found that it changes the pouring behavior of the powder such that dense and mechanically stable final materials are not readily achievable. Kronos 3025 (rutile, KRONOS Worldwide, Inc.) is found to consist of a broad range in size-distribution of the grains. The nanoparticles are washed out via sedimentation (\SI{1}{\kg} Kronos 3025 stirred in \SI{5}{\l} water, \SI{5}{\min} sedimentation time, \num{5} repetitions). The remaining particles are dried and sieved. Grains in the size range $\SI{20}{\um} < d < \SI{180}{\um}$ were selected for the compound material. For the final compounds, mass ratios of 4 : 1 $m_{\mathrm{Al}_2\mathrm{O}_3}$ : $m_{\mathrm{PA}12}$ and 14.7 : 1 $m_{\mathrm{TiO}_2}$ : $m_{\mathrm{PA}12}$ where used, as well as a negligible amount of carbon black to increase the absorption of the laser for melting ($\le$ 1g per 1000g compound).

For microwaves with wavelength in the \SI{}{\cm} range, the refractive index of the meta-material is the average of the refractive indices $n_i$ of the components $i$ weighted by their volume filling fractions $\phi_i$ and reads $n_{\text{eff}} = \sum_i \phi_i n_i$. However, since the compound material has inclusion of air, this does not give a reliable effective refractive index, and we measure it instead with a time of flight based measurement: the time difference $\Delta t$ between two microwave signals reflected at the front and at the back of the structure of length $L$ relates to the (average) refractive index $n_\mathrm{exp}$ of the structure via $n_\mathrm{exp} = c \Delta t / 2 L$. $\Delta t$ is obtained by measuring the peak distance of the \textsc{Fourier}-transformed S$_{11}$-signal of the vector network analyser.

\subsection{Measurement of transmission spectra}
\label{sec:method_macrospectroscopy}

In order to measure the transmittance through a photonic structure, a two port network is set up with a vector network analyzer (VNA, HP8719D, \SIrange{0.05}{13.5}{\GHz}). Two identical antennas (Aaronia PowerLOG 70180, \SIrange{0.7}{18}{\GHz}) emit (and receive) the signal and the sample is place in between both antennas in the wave-guide. The wave-guide of width \SI{10}{\a} and a length of $1~\mathrm{m}$ is used to reduce scattering of the electromagnetic wave in all directions beside the optical axis. It includes metallic shield acting as aperture of $1\mathrm{m}\times1\mathrm{m}$ at both sides of the wave-guide as sketched in (\autoref{fig:experimental_setup}), to avoid a short-cut of the electromagnetic wave bypassing the sample. The antennas emit electromagnetic waves of linear polarization which defines the $y$-axis and the wave-guide is oriented along the $z$-axis. To calculate the transmittance, the transmission coefficients with a sample in the guide is measured and squared to get the intensity. This intensity is normalized by the corresponding value of the spectrum without sample in the guide, to eliminate the specific features of the antenna and the wave-guide. To further average interference effects within the wave-guide, a linear stage is used that positions one antenna within a \SI{250}{\mm} range. The total transmittance is taken as the average measurements of transmission $S_{12} = S_{21} := t$ for various different positions of the emitting antenna. The spectra presented in \autoref{fig:experimental_diamond} and \autoref{fig:experimental_hpu_glass} are averaged about a hundred different positions. Based on the finit size of the antenna and using the wave-guide with shield, only a small fractions of the whole half space is detected. Thus, $R$ measures the \emph{directional reflection}.

\subsection{Refractive index}
\label{sec:ch3_timedomain}

By applying the \textsc{Fourier} transform to a reflection $S$-parameter ($s_{11}$ or $s_{22}$) as a function of frequency one can study its behavior in the time domain. The maximum frequency of the VNA dictates the time resolution and the frequency resolution dictates the maximum time that can be studied due to the inverse nature of the \textsc{Fourier} transform. In doing so, the impedances of the system can be evaluated. Any peak in the time domain indicates an impedance step like in time of flight measurements. Since the geometry of the setup and the sample is well known, the spectrally averaged refractive index $n_\text{eff}$ via $n_\text{eff} = \frac{c t}{2 L}$ can be extracted, where $L$ is given by the linear size of the sample and the corresponding peaks in the time domain are the discontinuities in impedance at the front and back side of the sample. The factor of $2$ arises since the signal runs back and forth to the VNA. Together with the volume fraction, $n_\text{eff}$ gives the high refractive index with the uncertainty $\delta n = n \left( \frac{\delta t}{t} + \frac{\delta L}{L}\right)$. This method provides a constant value for $n$ and dispersion is not accounted for. The refractive index of the compound material of PA12 with (Al$_2$O$_3$) was determined to be $n=1.83 \pm 0.35$ and that of PA12 with (TiO$_2$) to $2.15 \pm 0.38$. Since the uncertainty is large, the spectra are compared with numeric results of transmission spectra of the diamond lattice, where the refractive index was varied in steps of $\Delta n = 0.1$ (\autoref{sec:ch3_FDTD}). Best agreement was found for $n_{Al_2O_3} = 1.8$ and $n_{TiO_2}=2.1$.

\subsection{Simulations of transmission spectra} 
\label{sec:ch3_FDTD}

Finite difference time domain (FDTD) simulations with the ab-initio implementation of the software package MIT Electromagnetic Equation Propagation (\texttt{meep})~\cite{Oskooi2010} are performed in order to simulate the transmittance through a well defined structure.\\

The structure is represented as a spatially varying refractive index $n(\vec{r}) = \sqrt{\varepsilon(\vec{r})}$ relative to the refractive index of the vacuum $n=1$. Dimensionless length units denoted by \SI{1}{\a} are chosen. The structure of interest is sized to a cube with \SI{10}{\a} side length as in experimental investigations. A \SI{2}{\a} cladding of air is added in $z$-direction at both sides in which the source plane and flux plane lies which detects the passing intensity. Another \SI{2}{\a} cladding of absorbing material is added in $z$-direction at both sides. This ensures that radiation is not scattered back onto the sample. Metallic boundary conditions are chosen, resembling the experimental setup. The software then divides the geometry into a grid whereon the electromagnetic field is calculated for each timestep based on the previous fields strengths.\\

The source is set to emit a plane wave with a \textsc{Gauss}ian distribution in frequency. The reflection and transmission can be extracted as function of frequency by \textsc{Fourier} transforming the signal at the respective flux plane. The frequency is given in units of speed of light per length and denoted as \si{\cpa}. As for experimental spectroscopy methods, the data of a reference run without sample is done in order to normalize the data for each sample with the characteristic features of the source and the box. Loss is experienced only as finite time effect, since only the real part dielectric permittivity is used; due to the finite simulation time, radiation can still be trapped in the structure. Therefore the simulation has to run for a sufficiently long time and \num{1000} timesteps have proven to suffice. For low frequencies (long wavelengths) the simulations do not converge and results only above about \SI{0.1}{\cpa} are regarded as physically reasonable. At high frequencies (short wavelengths) the noise due to a finite resolution becomes larger and a compromise between finite computational expense and spacial accuracy needs to be found. A resolution in the range of \SIrange{15}{20}{\per\a} has proven to be a reasonable value. Note that the reflectance calculated this way includes all intensity that is redirected in general backward direction towards the source, including those scattered beside the Pointing-vector of the original plane wave. Thus, $R$ measures the \emph{hemispherical reflection}.

\section{Conclusion}

Three different photonic structures in the microwave range were constructed by 3D laser printing of a compound material. The structures of interest is a stealthy-hyperuniform one which was optimized to show an isotropic photonic band gap for sufficiently large refractive index. A diamond structure with well known but anisotropic band structure and a amorphous structure constructed from a 3D glass former were investigated as reference. For all three structures, the transmission spectra were measured in a wave-guide, using horn-antennas and a vector network analyser. The spectra were compared to finite difference time domain calculations (FDTD), where Maxwell's equations are solved numerically on a grid and to a priory band structure calculations.\\

The diamond structure shows a clear gap in the (100) and (111) direction, depending on orientation and increasing with refractive index, as expected. The stealthy hyperuniform structure shows a dip in transmission for $n_{exp}= 1.8\pm 0.35$ and an isotropic band gap for $n_{exp}= 2.1\pm 0.38$: the transmittance is reduced by three orders of magnitude and recovers almost two orders of magnitude back above the gap in the high frequency range. Thus it performs almost as well as the diamond along a symmetry axis. Furthermore the gap opens at a moderate refractive index - stealthy hyperuniformity is key to produce a gap for parameters more easily found in materials in the microwave or optical range.\\

The reduction in transmission of up to three orders of magnitude in the gap is true for rather small samples with ten nodes in linear dimensions. Doubling the structure in length will square the result. For both non-periodic structures, the hyperuniform and the glassy one, the transmittance does not recover $10^0$ in the high frequency range, neither in the experimental data where absorption can take places, nor in the simulations which lacks absorption in our model. This is attributed to the residual disorder of the structures which scatter intensity into the $4\pi$ space, not captured by the detector. An analysis of scattering by defects in the small wavelength spectrum in woodpile structures is given in \cite{Aeby2021}. While looking very similar by eye, the completely disordered glass structure does not show a gap at all, as expected.

%\appendix
%\section{}

\begin{acknowledgments}
We would like to thank Antonio Manuel Puertas, Almeria, Spain, who provided us with the datasets of hard sphere glasses. P.J.S and S.T. were supported  by the Army Research Office under Cooperative Agreement No. W911NF-22-2-0103. P.M.C acknowledges support from DOE under grant DE-SC0007991. L.S acknowledges financial support from the Department of Physics, University of Konstanz. P.K acknowledges support from Young Scholar Fund, University of Konstanz and G.M acknowledges SFB-1214 project B2 of the German Research Foundation. M.F. acknowledges EPSRC (United Kingdom) under  Strategic Equipment Grant No. EP/L02263X/1 (EP/M008576/1) and EPSRC (United Kingdom) under Grant No. EP/M027791/1 awards.
\end{acknowledgments}

\bibliography{references.bib}

\end{document}